\documentclass[12pt,preprint]{aastex}
 \newcommand{\WNM}      {\rm WNM}

\shorttitle{Warm Neutral Medium in Molecular Clouds?}
\shortauthors{Hennebelle \& Inutsuka}

\begin{document}
 
%
%
%
 
\title{Can Warm Neutral Medium Survive Inside Molecular Clouds?}

\author{Patrick Hennebelle\altaffilmark{1}
    and Shu-ichiro Inutsuka\altaffilmark{2} }
\altaffiltext{1}{ Laboratoire de radioastronomie millim{\'e}trique, 
                  UMR 8112 du CNRS, \\
             {\'E}cole normale sup{\'e}rieure et Observatoire de Paris,
              24 rue Lhomond, \\
              75231 Paris cedex 05, France;
              patrick.hennebelle@ens.fr}
\altaffiltext{2}{Department of Physics, Kyoto University, 
                 Kyoto 606-8502, Japan;
                 inutsuka@tap.scphys.kyoto-u.ac.jp}


\begin{abstract}
Recent high resolution numerical simulations have suggested that 
 the interstellar atomic hydrogen clouds have a complex two-phase 
 structure. 
Since molecular clouds form through the contraction of HI gas, 
 the question arises as to whether this structure is maintained 
 in the molecular phase or not.  
Here we investigate whether the warm neutral atomic hydrogen (WNM)
 can exist in molecular clouds.
We calculate how far a piece of WNM 
 which is not heated by the UV photons could penetrate into the cloud, 
 and find that in the absence of any heating 
 it is unlikely that large fraction of WNM survives 
 inside high pressure molecular clouds.
We then consider two possible heating mechanisms, namely 
 dissipation of turbulent energy and 
 dissipation of MHD waves propagating
 in the WNM inside the cloud. 
We find that the second one is sufficient to allow the 
 existence of WNM inside a molecular cloud of size $\simeq$ 1 pc 
 having pressure equal to $\simeq 10 \times P_{\rm ISM}$.
This result suggests the possibility that
  channels of magnetised WMN may provide efficient energy
  injection for sustaining internal turbulence which
  otherwise decays in a crossing time.
\end{abstract}


\maketitle

\section{Introduction}
There is strong observational evidence that the neutral  ISM consists of two distinct 
temperature regimes, the WNM and CNM (e.g. Heiles and Troland 2003).  
This is a natural result of most static 
 (e.g. Field et al. 1969, Wolfire et al. 1995), 
 time-dependent (Gerola et al. 1974), 
 and magnetohydrodynamical (e.g. Gazol et al. 2001; 
 de Avillez \& Breitschwerdt 2005; Inutsuka et al. 2005) 
 models of the ISM because of the temperature-dependence 
 of the interstellar cooling function.
Recent hydrodynamical numerical simulations with a high numerical resolution
approaching or resolving the Field length 
(Hennebelle \& P\'erault 1999, 2000; 
 Koyama \& Inutsuka 2000, 2002, 2004;
 Audit \& Hennebelle 2005; Heitsch et al. 2005; 
 V\'azquez-Semadeni et al. 2005; Hennebelle \& Passot 2006)  have shown 
that the structures of the interstellar atomic hydrogen is likely to be
 highly complex. In particular in these simulations, it is found that 
{\it i)} the structures of cold neutral medium (CNM)  are {\it locally}
 in near thermal pressure equilibrium with the 
surrounding WNM even if the flow is very dynamical on large scales,
{\it ii)} the 
CNM structures have a velocity dispersion close to the sound speed of the WNM
and  {\it iii)} the two phases appear to be highly interwoven. 
This picture of a dynamical two phase model  appears to be compatible with the observations 
supporting the existence of phases as well as the observations which finds 
 large fraction of thermally unstable gas (Heiles 2001) and large 
pressure fluctuations (Jenkins \& Jura 1983).

Since 
 molecular clouds are supposed to form 
 by  contraction of atomic gas, 
 the question of whether this complexity and the two-phase behaviour
 may persist even in molecular clouds arises immediately. 
We believe that
 the high turbulence which takes place in molecular clouds 
 reinforces this idea ; 
 since molecular clouds are surrounded by an HI halo, 
 the turbulence should almost inevitably induces some mixing between 
 HI and molecular gas.

From an observational point of view, we believe that the presence of warm
gas inside molecular gas is difficult  to prove or disprove 
unambiguously because 
of the surrounding HI halo.
However it is well established that molecular 
clouds are clumpy and have a low filling factor (see next section). 
Both facts are compatible with the idea that WNM may exist in those
clouds. 
Based on an analytic prediction of magnetic waves excited during 
a magnetic interaction of two magnetised clumps (Clifford \& Elmegreen 1983)
such two-phase model of molecular cloud has been considered by
   Falgarone \& Puget (1986) (see also  Elmegreen \& Combes 1992).
In their model, the giant molecular clouds are constituted of an ensemble 
of self-gravitating magnetised clumps which are surrounded by low pressure 
warm neutral gas which is heated either by ultraviolet photons or by 
the friction between ions and neutrals.

Here we investigate the possibility that  WNM may exist even inside dense
 molecular clumps at pressure significantly higher than the 
mean ISM pressure. 
For this purpose, we consider various mechanisms  and show the ranges of 
physical  conditions by order-of-magnitude calculations.  
In the second section we describe the physical processes
 which are important for the thermal equilibrium of WNM and 
 estimate how far a piece of WNM may penetrate into 
 a molecular cloud before it cools down.
In the third section, 
we  consider the  dissipation of MHD waves as a possible heating source 
of the WNM inside molecular clouds,   and we
 calculate the corresponding thermal equilibrium curve. We also discuss 
the effect of turbulence.
The final section concludes the paper.

\section{Basics}
\subsection{Notations and Assumptions}
We consider a molecular cloud of typical length $L$, mass $M$, 
volume $V$,
pressure $P$, density $\rho$, and velocity dispersion $\sigma$.
We assume that the cloud follows the Larson's law (Larson 1981), i.e., 
$M = 100 \, M_\odot \times (L/1 {\rm pc})^{2}$ and
\begin{equation}
\sigma \simeq \sigma^*  \times \sqrt{ {L \over 1 {\rm pc} }},
\label{sigma}
\end{equation}
where $\sigma^*=0.4$ km/s.
Thus the turbulent energy of the cloud is about $1/2 \times M \, \sigma ^2$.
The mean value of the number density is about
\begin{equation} 
   n             
 \simeq 570 \, {\rm cm}^{-3} \times 
                \left({ L \over 1 {\rm pc} } \right) ^{-1}.
 \label{density}
\end{equation} 

We assume that 
 two phases in near pressure equilibrium fill the cloud
 as in the atomic gas clouds; 
 one cold and dense phase 
 ($T_{\rm C}=10$ K, 
  $\rho_{\rm C} / m = n_{\rm C} > 10^3$ cm$^{-3}$ 
  where $m$ is the mean mass of the particle ),  
 and the other warm 
 and diffuse phase 
 ($T_{\WNM} \la 10^4$ K, 
  $n_{\WNM} = n_C \times T_{\rm C} / T_{\WNM} > 1$ cm$^{-3}$), 
 and we define the phase density contrast
 $r_\rho = n_{\rm C} / n _{\WNM} \la 1000$. 
If $f$ is the volume filling factor of the cold component, 
 then the volume occupied by the warm phase
 is $V_{\WNM} = (1-f) \times V$. 
With these definitions, we have
\begin{equation} 
  f   =     {( n - n_{\WNM} )\over( n_{\rm C} - n_{\WNM} ) }
    \approx {( n - n_{\WNM} )\over  n_{\rm C} }.
\end{equation} 
As mentioned in the introduction, 
 the simulations of turbulent atomic hydrogen clouds  
 have shown that the phases are deeply interwoven. 
If this structure is preserved during the transformation process
 from atomic gas to molecular gas, 
 channels of WNM gas permeating the molecular cloud should be naturally
 produced  and would remain as a natural outcome of the multiphase structure.

\begin{figure}
\includegraphics[width=8cm,angle=0]{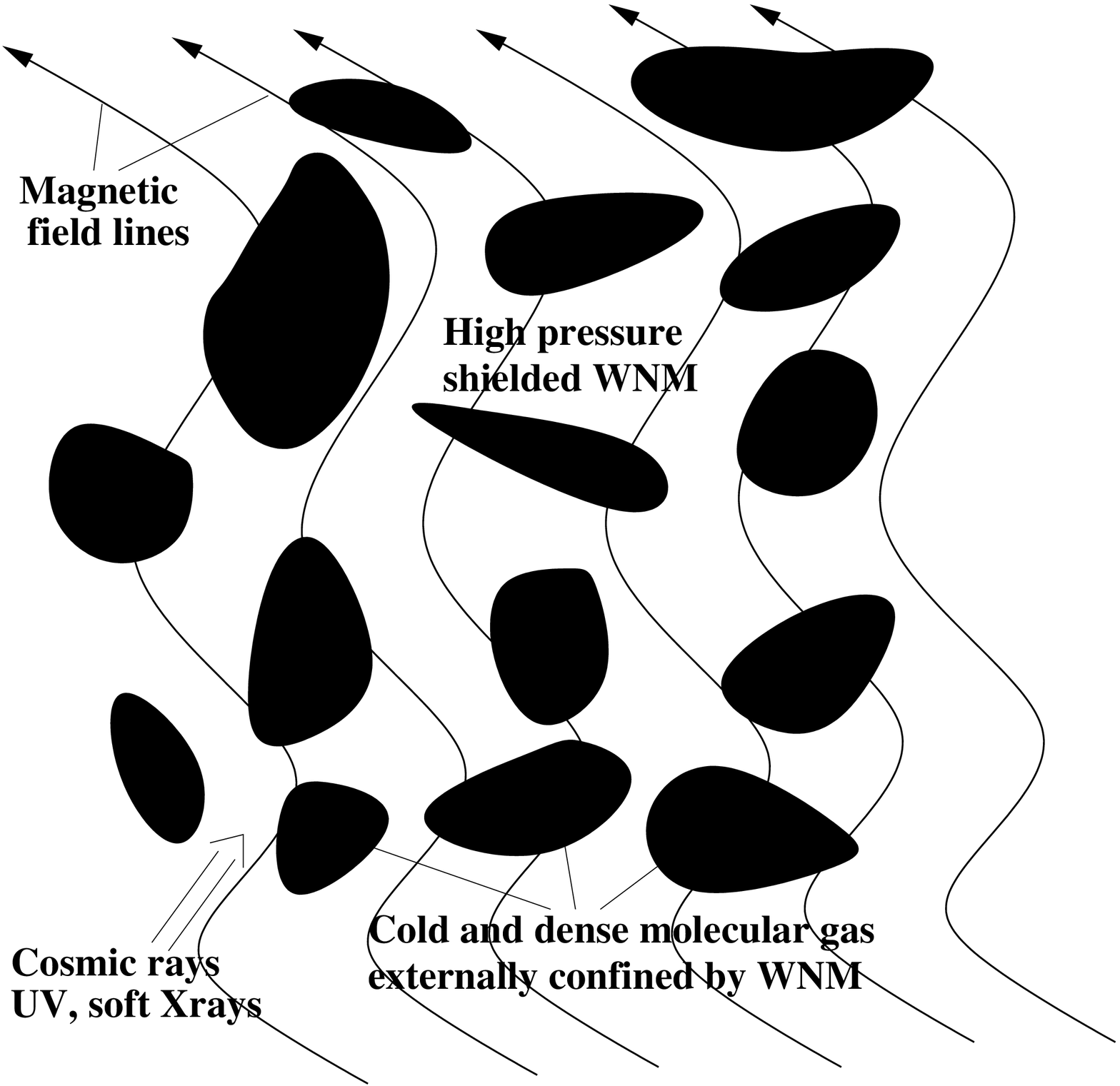}
\caption{Schematic picture illustrating the model of multiphase 
         magnetised molecular clouds.}
\label{picture}
\end{figure}

We will further assume that the whole cloud is permeated by a
 magnetic field $B$ having roughly the same intensity within the two phases.
In molecular clouds the observed average magnetic intensity  is about $20 \mu$G for 
$n=1000$ cm$^{-3}$ and for higher densities, B is roughly proportional 
to $\sqrt{n}$ (Crutcher 1999). 
As we discuss below, we will restrict our attention to clouds having a filling factor 
of about $\frac{1}{2}$ . 
In this case, the density of the cold phase is about $2 n$ and the 
 WNM density is $2 n / r_\rho \ga 2$ cm$^{-3}$.
Assuming that the mean magnetic intensity is nearly
 the same in the warm phase, 
 this indicates that the magnetic field intensity in WNM 
 embedded in two-phase molecular clouds
 is 
\begin{equation} 
 B_0 \simeq B^* \times \left( {n_{\WNM} \over 
 1 {\rm cm}^{-3} } \right)^{1/2}, 
 \label{magnetic}
\end{equation} 
where $B^* \la 20 / \sqrt{2} \simeq 14 \mu$G. 
Note that the assumption of magnetic intensity being 
roughly the same in the cold and the warm component is in good agreement 
with observations in the diffuse ISM (Troland \& Heiles 1986).
 The magnetic intensity  obtained by 
 direct measurements of the magnetic field in the diffuse ISM 
 indicates  values of 
 $5 \sim 6 \mu$G  in the WNM and in the CNM (Troland \& Heiles 1986, 
Heiles \& Troland 2006) where the mean density is respectively 
 about 0.5 cm$^{-3}$ and 50 cm$^{-3}$. These measurements suggest 
 values of $B^*$ smaller than 15 $\mu$G which may indicate  lower filling factor
 or that our assumption of uniform magnetic field in the cloud does not hold exactly.
In the following, we adopt $B ^* \simeq 10 \mu$G 
 that seems in better agreement with the magnetic intensities 
 measured in the diffuse ISM.
We also assume that the fluctuating component of the magnetic field, $\sqrt{\langle \delta B^2 \rangle}$,  
is equal to the mean magnetic field, $B_0=\langle B \rangle$ which corresponds to 
equipartition between the magnetic energy of the mean and the fluctuating 
part of the magnetic field. In the following we refer 
to the energy of the fluctuating component of the magnetic field 
within the WNM as the magnetic wave energy.
Figure \ref{picture} shows a schematic picture 
 that illustrates the model.

Finally we also assume that the cloud is embedded in a diffuse gas 
 of standard WNM having a density 
 $n_{\rm WNM} ^{\rm ext} \simeq 0.5$ cm$^{-3}$, 
 a magnetic field of uniform component 
 $B_{\rm ext} \simeq 6 \mu$G, 
 and a fluctuating component 
 $\sqrt{\langle \delta B^2_{\rm ext} \rangle} \simeq B_{\rm ext}$.
The cloud therefore receives a flux of magnetic energy from the external
interstellar medium.
This magnetic energy is produced at large scales 
 through various mechanisms, 
 such as clump magnetic interaction 
 (Clifford \& Elmegreen 1983, Falgarone \& Puget 1986), 
 supernovae explosions or galactic differential rotation 
 (e.g. MacLow \& Klessen 2004).

According to equations (2) and (3), 
 the filling factor of the cold molecular component in a 10-pc cloud 
 is  very small (say, less than 0.1). 
It is therefore very likely that in such a
cloud, the external radiation is not very different in most of the cloud volume,
 from its value outside the cloud. 
Therefore the conditions are likely to be 
 similar to the standard ISM considered by Wolfire et al. (1995) 
 so that WNM can exist inside such a large complex 
 if the pressure in the rarefied area is comparable 
 to the ISM pressure.
On the other hand, for a 1 pc cloud, either $f \simeq 0.5$ or 
 $n_c$ is higher than $10^3$ cm$^{-3}$ indicating 
 a high thermal pressure. 
In both cases the UV background is not intense 
 enough to heat  the warm phase (see Section \ref{cosmic}).
Therefore in about one cooling time, 
 WNM cools down into cold gas  unless it is
 heated by another source of energy. 
In the following section we describe the relevant physical phenomena 
(degree of ionization and recombination time,
heating due to cosmic rays, UV and soft X-rays, cooling rate and cooling time) and
investigate how far a piece of WNM which is not heated could penetrate  inside molecular clouds.

\subsection{Cosmic Rays, UV, and Soft X-Rays}
\label{cosmic}
Cosmic rays, far ultraviolet (FUV) photons and soft X-rays are the main  
 ionization  sources of the WNM. They also constitute the main heating 
sources of the standard WNM.

The value of the ionization rate due to cosmic rays is not well 
 constrained and could be possibly very inhomogeneous. 
In molecular clouds, cosmic ray total 
(primary and secondary)
 ionization rate of  $\zeta=2 \sim 7 \times 10^{-17}$ s$^{-1}$ 
(e.g. Goldsmith 2001,  Wolfire et al. 1995) are usually considered. 
However in diffuse clouds, 
 values 40 times higher have been proposed by McCall et al. (2003)
 for explaining the high H$_3^+$ abundance 
 (see also, Le Petit et al. 2004). 
Recently Padoan \& Scalo (2004) proposed that 
 confinement of cosmic rays by  self-generated 
 MHD waves may naturally accounts for these large variations.
Since no measurement of cosmic ray ionization rate in the WNM is available 
 in the litterature and 
 since we are proposing a model in which 
 cold molecular gas is embedded in warm HI medium, 
 the cosmic ray ionization rate relevant for our study is highly uncertain. 
In the following, we therefore consider a standard ionization rate of 
$\zeta=3 \times 10^{-17}$ s$^{-1}$ and discuss the consequences of higher values.
The corresponding heating rate  is $10^{-27}$ erg s$^{-1}$ (Goldsmith 2001).
As we will see later this is too small to contribute 
 significantly to the heating of WNM.

As shown in Wolfire et al. (1995), the photoelectric heating 
 from small grains and PAHs due to FUV photons is the most important
 source of heating of the standard WNM and is about  
 $1.0 \times 10^{-24} \epsilon G_0$ erg s$^{-1}$, 
 where $\epsilon$ is the photoelectric efficiency and 
 is equal to about $\le$0.1 and $G_0$
 is the incident FUV field normalized to Habing (1968)'s estimate 
 of the local interstellar value. 
In the case of the standard ISM this heating appears to be sufficient to heat WNM 
of density up to 1 cm$^{-3}$. 
In the case of molecular clouds, the UV field is 
 reduced by dust extinction by a factor 
 $\simeq \exp (-1.8 \times A_v)$ where $A_v$ is the extinction in visual 
magnitude. 
For a 1-pc cloud of average gas density 10$^3$ cm$^{-3}$, 
 this gives roughly $A _v \simeq 1$ reducing 
 the incident FUV flux by a factor $\simeq 6$.
Note that since in our model the cloud is a multiphase   and clumpy object, 
 the FUV penetration should be enhanced by the diffusion processes 
 (Boiss\'e 1990) and the incident FUV flux larger than this value. 
In any case, however, the heating  due to standard FUV field is not sufficient 
 to permit the  existence of WNM at pressure larger than the standard ISM pressure
and  therefore cannot permit the existence of WNM inside most molecular clouds.
Thus in this paper we assume, for simplicity,  
 that the heating due to UV radiation field is negligible. 
Note that this approximation 
 corresponds to the most pessimistic assumption regarding 
 the existence of WNM inside molecular clouds, and may not be appropriate
 for a low pressure isolated molecular cloud or 
 for a cloud in the neighborhood of a strong UV source.

As shown in Wolfire et al. (1995) the ionization and heating due to 
 soft X-rays becomes negligible for column densities larger than 
 $10^{20}$ cm$^{-2}$.
Such a column density corresponds to a length of less than  0.03 pc 
for the cold 
component which we assume surrounds the warm phase that we are considering.

\subsection{Ionization and Cooling Rate}
The recombination rate of electron onto proton is about
 $2.6 \times 10^{-13} T_4 ^{-0.7} n_e$
 s$^{-1}$, where $T_4= T /10000 K$ and $n_e$ is the electron densities.
This leads to an ionization fraction of 
\begin{eqnarray} 
x \equiv n_e/n_{\WNM} \simeq 10^{-2}  \left( {\zeta \over 3 \times 10^{-17} s^{-1} } \right) ^{1/2}
 \left( {n _{\WNM} \over 1  {\rm cm}^{-3} } \right)^{ - 1/2} \left( { T \over 8000 K } \right) ^{0.35} ,
\label{ionisation}
\end{eqnarray}
 which is about 10 times smaller than in the standard WNM. 
The other possible contributions are ionization of heavy elements 
 (mainly carbon) due to FUV and 
 ionization of atomic hydrogen due to soft X-rays.
However, the former is 
 negligible since the carbon abundance is about 
 $3 \times 10^{-4}$ and
 the latter is also negligible as explained in Section ~\ref{cosmic}. 
Before the piece of WNM penetrates into the cloud, 
 its ionization fraction is the standard value of the ISM, 
 i.e., $x \simeq 0.1$. 
Once it enters inside the cloud, it takes about one recombination time, 
 $\tau _{\rm rec} \simeq 1$Myr to reach  the equilibrium value 
 within the molecular cloud, i.e  $x \simeq 10^{-2}$.


We have calculated the cooling rate by considering the standard cooling 
mechanisms of the atomic gas that are 
 described in Wolfire et al. (1995), namely emission of Lyman $\alpha$, 
 metastable lines (CII, OI) and fine-structure lines (CII, OI) taking 
 the values from Hollenbach \& McKee (1989) and Wolfire et al. (1995) 
with  heavy elements abundances  corresponding to the 
solar neighborhood.
Note that the ionization degree of  carbon is very uncertain since
as discussed in the previous section the UV flux is difficult to estimate.
Therefore we make the assumption that  carbon is fully ionized. This 
again corresponds to the most pessimistic assumption regarding existence 
of WNM inside molecular clouds since lower 
value of C$^+$ abundances  will reduce the cooling (by about 20\%). 
Note also that in spite of the fact that the whole cloud 
 we are considering has a high opacity, 
 the WNM embedded inside this cloud has 
 a low column density and remains optically thin. 
Thus, the Lyman alpha photons emitted in the 
WNM propagate  until they are absorbed by the dust into a
 piece of cold and dense gas.
  The corresponding energy is then reradiated  by the cold component 
 and finally leaves the cloud.
Opacity effects are negligible for the [OI] 63 $\mu$m photons 
emitted by the WNM and these photons  escape the cloud. 
For the [CII] 158 $\mu$m photons, the   opacity effects may depend on the 
exact structure of the cloud but remain modest (Tielens \& Hollenbach
1985).

The main difference from the  standard ISM calculations 
  is about 10 times smaller ionization degree of the WNM. 
Therefore the radiative cooling is a few times smaller as well 
 (see e.g. Dalgarno \& McCray 1972), leading to a
 cooling rate of about $4 \times 10^{-26}$ erg cm$^{3}$  s$^{-1}$ at 
 $T=8000$ K. 

\begin{figure}
\includegraphics[width=8cm,angle=0]{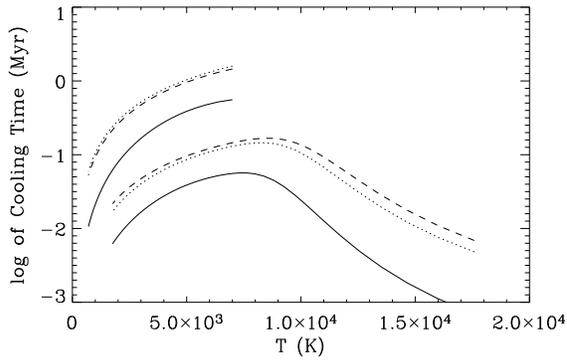}
\caption{Cooling time of the WNM when it enters a molecular cloud
of pressure equal to the ISM pressure, $P_{\rm ISM}$ (3 top curves), 
 and $10 \times P_{\rm ISM}$ (3 bottom curves).
Full line is for an ionization degree, $x$, of 0.1, dotted line is for 
$x=$0.01 and 
dashed line is when the ionization equilibrium is assumed.}
\label{cooling_time}
\end{figure}

Figure \ref{cooling_time} shows the cooling time of the WNM when it 
 enters into the cloud as a function of temperature. 
The  density is
 obtained by assuming pressure equilibrium. The largest temperature 
 corresponds to the case of a piece of WNM 
 which  has been 
 suddenly compressed (and therefore heated) to the cloud pressure.
Because of the cooling and the absence of  heating,  the
 temperature of the fluid element decreases monotonically with time, 
 and the relevant cooling timescale is given by the largest value.

The cooling time strongly depends on the ionization degree.
However since it is always smaller than about 1 Myr and since
 the recombination time is about $1$ Myr, the ionization degree 
 is likely to be given by the full line 
 (except maybe for  $P= P_{\rm ISM}$).

\subsection{Penetration Length of WNM in Absence of Heating}
As discussed previously, heating due to FUV and soft X-rays is not efficient
 enough to provide a significant heating deep inside molecular clouds. 
We therefore 
calculate how deep the piece of WNM may penetrate inside the cloud
without being heated. The corresponding length is given by the product of 
the WNM velocity by the cooling time.

In numerical simulations of HI flows (Koyama \& Inutsuka 2002, 
Audit \& Hennebelle 2005), it is found that the WNM velocity is higher by a 
factor of a few than the velocity dispersion of the CNM. 
We therefore expect  that 
$u _{\WNM} \simeq \alpha _{cross} \times \sigma _c$,
where $\alpha _{cross}$ is a factor of a few. 
This is consistent with the idea that the cloud is continuously 
 swept up by the surrounding HI gas which has an internal velocity 
 dispersion equal to a fraction of the WNM sound speed, 
 $C _{\WNM} \simeq 10$ km/s.

Assuming that the WNM velocity inside the cloud is about 1 km/s
 (note that for 1-pc cloud, the Larson's law gives 
  a velocity dispersion of about 0.4 km/s), 
 we find  with Figure \ref{cooling_time} that 
the piece of WNM cannot penetrate into the cloud deeper than 
 $0.5 \sim 1$ pc for $P= P_{\rm ISM}$ and 
 0.1 pc for $P= 10 \times P_{\rm ISM}$. 
Thus we must conclude that in the absence of heating 
 WNM can exist inside low pressure molecular clouds 
 ($P \simeq P_{\rm ISM}$) of size $L \simeq 1$ pc but cannot 
 penetrate significantly inside high pressure molecular clouds 
 of size larger than 0.1 pc.

\section{Heating Rate due to Mechanical Energy Dissipation}
Here we estimate the heating rate of the warm neutral phase
due to mechanical energy dissipation  in the cloud.
First we consider the dissipation of magnetic waves 
 that propagate into the warm phase inside the molecular cloud. 
Similar calculations have been performed by Scalo (1977) 
in the case of dense molecular gas  and 
by Ferri\`ere et al. (1988)  in the case of standard WNM.
Both found that the dissipation of magnetic waves can provide 
a substantial heating rate.
We then estimate the amount of energy
available from the  dissipation of turbulent motions 
 which are observed in the dense gas of molecular clouds.

\subsection{Dissipation of Magnetohydrodynamical Waves in WNM}

\subsubsection{Wave Dissipation and Wave Energy}

The magnetic energy of the waves per unit  volume is 
 $\langle\delta B ^2\rangle / 8 \pi = (\langle\delta B^2\rangle / B_0^2)  B_0 ^2/ 8 \pi$, 
 where $B_0$ is the mean magnetic field, and $\delta B$ is 
 the fluctuating part of the magnetic field. 
In the following we will  assume 
 $B_0^2 \sim \langle\delta B^2\rangle$ 
 which corresponds to energy equipartition 
 between the magnetic energies of the mean field and 
 the fluctuating component.
Assuming that  the kinematic energy of the waves 
 is comparable to the magnetic energy 
 (as is the case for pure Alfv\'en waves),
 the wave energy, 
\begin{equation}
 E_{\rm wave} = {\langle\delta B ^2 \rangle \over B_0^2} 
                 { B_0^2 \over 4 \pi }
\label{Ewave}
\end{equation} 
and
 the total amount of energy per particles of WNM available 
 from the waves is therefore, 
 $ V_{\WNM} / N_{\WNM} \times 
     \langle\delta B ^2\rangle / 4 \pi  \simeq  (\langle\delta B^2\rangle / B_0^2)  B_0^2 / 4 \pi / n_{\WNM}$ where $N_{\WNM}$ is the number of WNM particles within the cloud.

The propagation of the Alfv\'en waves in a weakly ionized gas 
 has been studied by Kulsrud \& Pearce (1969).   
They show that for a wavelength smaller than the critical wavelength, 
 \begin{equation}
       \lambda_{{\rm crit}} = 
                { \pi v_{\rm A}  \over \gamma _{\rm da}  \rho_{\rm i}}, 
                                                    \label{crit1}
 \end{equation} 
 the wave does not propagate, 
 whereas for a wavelength larger than the critical wavelength, 
 the wave propagates but dissipates in a timescale given by 
 \begin{equation}
       t _{\rm da} =  {2  \gamma _{\rm da}  \rho_{\rm i}  
                   \over  v_{\rm A}^2 (2 \pi/\lambda)^2}, 
 \end{equation}
  where 
  $\lambda$ is the wavelength, 
  $\gamma _{\rm da}$ is the friction coefficient between ions 
  and neutrals, 
  $\rho_{\rm n}$ and $\rho_{\rm i}$ are the neutral and ion densities,
  and $v_{\rm A}$ is the Alfv\'en velocity. 
The expression of $\gamma _{\rm da}$ for a weakly ionized hydrogen gas
 has been derived by various authors 
 (e.g. Osterbrock 1961, Draine 1980, Mouschovias \& Paleologou 1981).  
Here we adopt the most recent value obtained by 
 Glassgold et al. (2005), 
 $\gamma_{\rm da} = 5.7 \times 10^{14}$ cm$^3$ s$^{-1}$ g$^{-1} 
  (v _{\rm rms} / {\rm km s}^{-1})^{0.75}$ 
 where $v_{\rm rms} = \sqrt{8 k T / \pi m}$ 
 is the mean thermal speed. 
This leads to 
 $\gamma _{\rm da} \simeq 
  3.4 \times 10 ^{15}$ cm$^3$ g$^{-1}$ s$^{-1} 
  (T/8000 {\rm K})^{0.375}$ .

Using the expression of the magnetic field (eq. [\ref{magnetic}]) 
 and the ionization rate (eq.[\ref{ionisation}]),  
 we obtain
 \begin{equation}
        \lambda _{\rm crit}  \simeq 
            0.024 \, {\rm pc}
        \left( {  n_{\WNM} \over 1 \, {\rm cm}^{-3} } 
        \right) ^{-1/2}
        \left( {T \over 8000 {\rm K} } \right)^{-0.725}
        \left( \zeta \over 3 \cdot 10^{-17} {\rm s}^{-1} \right)^{-1/2}
        \left( {B^* \over 10 \, {\rm \mu G} } \right). 
        \label{lcrit}
 \end{equation}

\subsubsection{Replenishment of Mechanical Energy}
Another important issue is on the mechanism for maintaining 
 turbulent motions in molecular clouds. 
In the model that we are investigating, 
 the WNM is heated by the dissipation of MHD waves. 
Therefore these waves must be continuously replenished 
 from the outside, otherwise the heating source will 
 disappear in the dissipation timescale. 
In the case of a single phase molecular cloud, 
 Nakano (1998) pointed out that the dissipation timescale
 is shorter than the crossing time, 
 which implies that internal turbulence cannot be easily 
 driven from the outside.
Here we examine three conditions that must be satisfied in order to 
 enable energy replenishment from the outside. 

The first condition is that the timescale for the waves traveling 
 into the WNM to reach  the cloud centre, 
 $t_{\rm cross} = (L/2) / v_{\rm A}$, must be smaller
 than or comparable to their dissipation timescale. 
The ratio between crossing timescale and dissipation timescale 
 is given by
 \begin{eqnarray}
  && { t_{\rm cross} \over t_{\rm da} } = 
     {  \pi ^2 v_{\rm A} L \over 
       \lambda^2 \gamma_{\rm da} \rho_{\rm i} } 
     = \frac{\pi \lambda_{\rm crit} L}{\lambda^2} . 
    \label{crit2}
 \end{eqnarray}
Thus, the condition $t_{\rm cross}/t_{\rm da} \le 1$ leads to
 $\lambda \ge \lambda_{\rm cross}$ where   
 \begin{eqnarray}
  \lefteqn{ \lambda _{\rm cross} \equiv 
           \left( \pi L \lambda_{\rm crit} \right)^{1/2}  } 
                                                \label{lcross} \\ 
  && \simeq  0.27 \, {\rm pc}
  \left( {  n_{\WNM} \over 1 \, {\rm cm}^{-3} } \right) ^{-1/4}
  \left( {T \over 8000 {\rm K} } \right)^{-0.362}
  \left( \zeta \over 3 \cdot 10^{-17} {\rm s}^{-1} \right)^{-1/4}
  \left( {L \over 1 \, {\rm pc}} \right) ^{1/2}
  \left( {B^* \over 10 \, {\rm \mu G} } \right) ^{1/2}. 
  \nonumber 
 \end{eqnarray}
Note that this expression is valid only if $L \ge \lambda_{\rm cross}$.
Therefore the smallest value of $\lambda_{\rm cross}$ is obtained when 
$L \simeq \lambda_{\rm cross}$ and is about $\pi \lambda_{\rm crit}$.

If the cloud is very anisotropic
then the length of the shortest axis should be used in the expression of $\lambda _{\rm cross}$.

\begin{figure}
\includegraphics[width=8cm,angle=0]{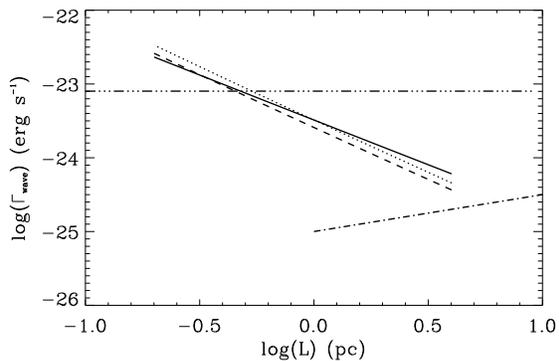}
\caption{Heating rates due to MHD waves dissipation
 $\Gamma_{\rm wave}$ 
stated by equation~(\ref{gam_gen}). 
  Full lines are for clouds following Larson's laws so that 
$n \propto 1/L$. Dotted and dashed lines are for clouds of gas densities
1 and 3 cm$^{-3}$ respectively. The dot-dashed  line displays the heating due 
to the dissipation of turbulence. The triple dot-dashed  line displays the 
 heating that would be obtained by dissipating all the available magnetic 
wave energy in standard ISM conditions (see eq.~[\ref{Gamma_lim}]) for clouds following
the Larson's laws. }
\label{thermax}
\end{figure}

\begin{figure}
\includegraphics[width=8cm,angle=0]{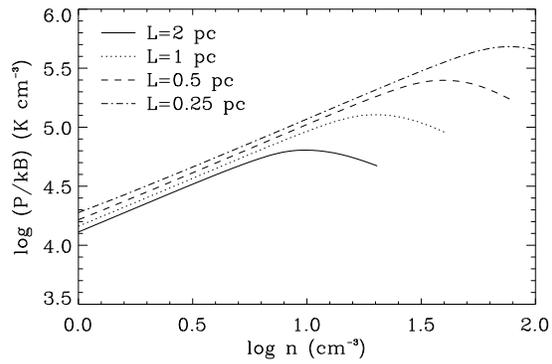}
\caption{Thermal equilibrium curves for a heating rate due to 
MHD wave dissipation $\Gamma _{\rm wave}$ and
for four cloud sizes. Only densities corresponding to the warm phase
are shown. 
The  points which have positive slope correspond to thermally stable state
whereas the one having negative slope are unstable.  The existence of WNM
requires that the pressure is lower than the largest pressure reached by 
the thermal equilibrium curve.}
\label{thermal_equilibrium}
\end{figure}

The second condition is that the energy flux due to MHD waves 
which penetrate inside the cloud must be larger than the total energy
dissipated by unit of time inside the cloud. This flux of energy 
cannot be easily computed but the maximum value can be estimated as 
  $F _{E}\simeq 4\pi (1-f)^{2/3} (L/2)^2 v_{\rm A}^{\rm ext} E_{\rm wave}^{\rm ext}$ 
where 
$v_{\rm A} ^{\rm ext}$ and  $E_{\rm wave} ^{\rm ext}$ are respectively the Alfv\'en velocity and 
the total energy of the waves outside the cloud and $4 \pi  (L/2)^2$ is the surface of
the cloud. 
The factor $(1-f)^{2/3}$ takes into account the fact that 
 only a fraction of the surface is constituted by WNM. 
The number of WNM particles is 
   $V    (1-f) \, n_{\rm WNM}$,  
 therefore one must have 
 $\Gamma _{\rm wave} V _c (1-f) \, n_{\rm WNM} \le  F_{E}$ 
 which leads to
\begin{eqnarray} 
  \Gamma _{\rm wave}   \le \Gamma _{\rm wave} ^{\rm lim} \equiv  
  {6 v_{\rm A} ^{\rm ext} E_{\rm wave} ^{\rm ext} \over 
  n_{\rm WNM} L} =  {6 B_{\rm ext} \langle\delta B_{\rm ext}^2\rangle 
  \over 
  (4 \pi)^{3/2} \sqrt{m} n_{\rm WNM} \sqrt{n_{\rm WNM} ^{\rm ext}} L },  
  \label{Gamma_lim}
\end{eqnarray} 
 where $E_{\rm wave} ^{\rm ext} = \delta B_{\rm ext}^2 / 4 \pi $ 
 and where the factor $(1-f)^{1/3}$ has been dropped for simplicity 
 since it is on the order of unity 
 unless the value of $f$ is close to unity. 
The external gas density and magnetic intensity outside the cloud 
 presumably vary significantly from place to place, 
 in  particular if the cloud forms dynamically.
To estimate the largest
heating which can be possibly due to  input of external energy we 
adopt  standard ISM conditions, $B _{\rm ext} \simeq 6 \mu$G,
$\langle\delta B_{\rm ext}^2\rangle \simeq  B_{\rm ext}^2$ and
 $n_{\rm WNM}^{\rm ext} \simeq 0.5$ cm$^{-3}$ which leads to
\begin{eqnarray} 
  \Gamma _{\rm wave} ^{\rm lim}   \simeq  8 \times 10^{-24} \,{\rm erg \, s}^{-1}
\left( { \langle\delta B_{\rm ext}^2\rangle \over  B_{\rm ext} ^2 } \right) 
  \left( { B_{\rm ext} \over 6 \, \mu {\rm G} } \right) ^3 
\left( {n_{\rm WNM}    \over 1 \, {\rm cm}^{-3}} \right)  ^{-1}  
\left( {n_{\rm WNM}   ^{\rm ext} \over 0.5 \, {\rm cm}^{-3}} \right)^{-1/2}  
 \left( {L    \over 1 \, {\rm pc}} \right)  ^{-1}  ,
\end{eqnarray} 
For  clouds which follows the Larson's law (eq.~[\ref{density}]), 
one finds that $(n_{\rm WNM} / 1 \, {\rm cm}^{-3}) (L/1 \, {\rm pc}) \simeq 1$, 
so that $\Gamma _{\rm wave}  ^{\rm lim}  \simeq  8 \times 10^{-24}$ ${\rm erg \, s}^{-1}$
is independent of gas density and cloud length.
As we will see in following sections (see Fig. \ref{thermax} and 
 \ref{thermal_equilibrium}), 
 this is sufficient to heat WNM inside molecular cloud 
 of size $L \ge 0.5$ pc at the largest equilibrium pressures 
 that are computed below.

The third condition is related to the existence of the channels of 
 magnetised warm gas which must permeate the molecular cloud in order 
for the energy to be replenished from the outside. 
Because of heat conduction 
 the tunnels of WNM cannot be infinitely thin and 
 in any case should be small compared to the size of the cloud. 
The smallest size of a piece of WNM embedded into cold gas 
 is given by the Field length of WNM which is the typical size 
 of a thermal front between the two phases. 
For smaller size the heat flux due to thermal conductivity 
 between the two phases will cool the WNM.
The Field length is about
\begin{equation}
\lambda_F \simeq \sqrt{ {M_p \kappa(T) T \over \Gamma }  }, 
\label{field}
\end{equation}
where $\kappa(T)$ is the thermal conductivity, $\kappa(T) = 5/3 C_v \eta(T)$ and 
$\eta = 5.7 \times 10^{-5} (T/ 1 K)^{1/2}$ g cm$^{-1}$ s$^{-1}$ and 
$\Gamma$ is the heating rate.  In the standard WNM, the Field length is about 0.1 pc. 
As we will see in the next section, for the case that we are studying the 
heating rate can be more than 10 times higher than the heating rate of the WNM in standard ISM 
(see section 2.2) leading to a Field length smaller than $0.1 / \sqrt{10} = 0.03$ pc.
The value is therefore small compared to the sizes of molecular clouds 
 that enable the existence of the channels.
Finally we also estimate the size of the smallest channels that can 
 exist in spite of the thermal diffusivity during one dynamical time. 
If $\lambda_{\rm mfp}$ is the mean free path, then a particle undergoes 
a collision in a time of about $\lambda_{\rm mfp} / C_{\rm s}$.
During the collision with a colder particles, 
 the WNM particle losses its thermal energy.   
Therefore the time required in order for the whole finger of WNM 
 to cool down by collision with colder particles is 
 $(R/\lambda_{\rm mfp})^2 \times \lambda_{\rm mfp} / C_{\rm s}=
 R^2/ \lambda_{\rm mfp} C_{\rm s}$ where $R$ is the radius of the finger. 
If we require that the finger exists during 
a cloud crossing time, $L / C_{\rm s}$, we find that 
$R \simeq \sqrt{L \lambda_{\rm mfp}}$. For a density of 
$n \simeq 1$ cm$^{-3}$, $\lambda_{\rm mfp} \simeq 10^{16}$ cm. Therefore 
$R \simeq 0.03$ pc for $L = 1$ pc which is thus  comparable of the 
Field length within the cloud.

\subsubsection{Heating Rate}
Here we calculate the heating rate of the WNM 
 due to MHD wave dissipation.
Since $t_{\rm da}$ depends on $\lambda$, 
 a wave powerspectrum has to be considered.

We assume that the (isotropic) power spectrum of MHD waves 
 ${\cal E}(k)$ has significant values only between 
 $k_{\rm min}=2\pi/\lambda_{\rm max}$ and 
 $k_{\rm max}=2\pi/\lambda_{\rm min}$, 
 thus, the wave energy 
$
 E_{\rm wave} = \int_{k_{\rm min}}^{k_{\rm max}} {\cal E} (k) dk,  
$
 where we simply require  
\begin{equation}
  \lambda_{\rm crit} \le \lambda_{\rm min} 
  \le \lambda_{\rm max} \la L.  
\end{equation}
The value of $\lambda_{\rm min}$ is uncertain.
On one hand, equation~(\ref{lcross}) gives the smallest wavelength 
 which can be injected from the outside. 
On the other hand, 
 wave steepening of long wavelength waves could 
 generate waves of wavelengths close to $\lambda _{\rm crit}$.
For the sake of simplicity we ignore this last possibility and 
 assume that 
 $k_{\rm max} = 2 \pi/\lambda_{\rm cross}$. As we discuss below, larger 
values of $k_{\rm max}$ lead to larger heating.

Since power spectra are often observed to be power-law, 
 we further assume ${\cal E}(k) = {\cal E}_0 (k_{\rm min}/k)^p$
 where we may assume $1 < p < 2$ according to the observations of 
 molecular clouds. 
Note that the so-called Kolmogorov spectrum corresponds to $p=5/3$. 
We have
 \begin{equation}
  E_{\rm wave} = 
  \int_{k_{\rm min}}^{k_{\rm max}} {\cal E}_0 (k_{\rm min}/k)^p dk
  = { {\cal E}_0 k_{\rm min} \over p -1  } 
    \left( 1 - \left(  {k_{\rm min} \over  k _{\rm max}} \right)^{p-1} 
    \right).
  \label{spectrum}  
 \end{equation}
Obviously the energy is dominated by the waves of the largest 
 wavelengths if $p > 1$.

The energy dissipation rate (per volume and time) is
 \begin{equation}
 \dot{E}_{\rm wave} = \int_{k_{\rm min}}^{k_{\rm max}} 
                            \frac{{\cal E}(k)}{t_{\rm da}(k)} dk
 = \int_{k_{\rm min}}^{k_{\rm max}}  
   {\cal E}_0 (k_{\rm min}/k)^p {v_{\rm A}^2 \over 2 \rho_{\rm i} 
  \gamma_ {\rm da}} k^2 dk.
 \end{equation}
Strictly  speaking, $v_{\rm A}$ depends on $k$. 
 However taking this into account would add 
 more complexity to the final result without improving 
 significantly its physical meaning.  
 We therefore ignore this and write with equation (\ref{spectrum}) and
using the expression for $t_{\rm da}$ 
\begin{equation}
 \dot{E}_{\rm wave} = {\cal E}_0 {v_{\rm A}^2 \over 2 \rho_{\rm i} 
 \gamma_ {\rm da}}
 \int_{k_{\rm min}}^{k_{\rm max}}   (k_{\rm min}/k)^p  k^2 dk =
 {p-1 \over 3-p} \times 
 {E _{\rm wave}  v_{\rm A}^2 k_{\rm min}^2 \over 2 \rho_{\rm i} \gamma_ {\rm da} }  
 { (k_{\rm max}/ k_{\rm min})^{3-p} -1  \over 
  1 - (k_{\rm min}/ k_{\rm max})^{p-1}}.
 \end{equation}
Therefore in the case of the wave powerspectrum, 
 the heating rate due to MHD wave 
 dissipation can be written as 
 \begin{equation}
  \Gamma_{\rm wave} = \dot{E}_{\rm wave} = 
  f(p, \frac{k_{\rm max}}{k_{\rm min}}) 
  \Gamma_{\rm wave,0} (\lambda_{\rm max}) 
                                        \label{gam_gen}
 \end{equation}
 where 
 \begin{equation}
  f(p,q) = {p-1 \over 3-p} \times
  { q^{3-p} -1 \over 1 - q^{1-p}},  \label{eq:f}
 \end{equation}
 \begin{eqnarray}
   \Gamma_ {\rm wave,0} (\lambda) =  
   { E_{\rm wave} \over  n _{\WNM} t_{da} } = 
   {B_0^2  \langle\delta B ^2\rangle \over 8  \gamma _{da}
    n _{\WNM} ^3 m_{\rm i} m x   \lambda^2 }.
 \end{eqnarray}
$\Gamma _{\rm wave,0} (\lambda)$ is the heating that would 
 be obtained with monochromatic waves of wavelength equal to 
 $\lambda$. 
Taking into account the expression of the magnetic field and 
 the ionization rate, 
 this leads to
 \begin{eqnarray}
  \Gamma_{\rm wave,0} (\lambda)
  &\simeq& 8 \times 10^{-25} {\rm erg \; s^{-1}}
 \\ \nonumber
  && \left( {  n_{\WNM} \over 1 \, {\rm cm}^{-3} } \right) ^{-1/2}
  \left( {T \over 8000 {\rm K} } \right)^{-0.725}
  \left( \zeta \over 3 \cdot 10^{-17} {\rm s}^{-1} \right)^{-1/2}
  \left( {\lambda \over 1 \, {\rm pc}} \right) ^{-2} 
  \left( {B^* \over 10 \, {\rm \mu G} } \right)^4
  \left( {\langle\delta B^2\rangle \over B ^2} \right).
    \label{gamwave}
 \end{eqnarray}
It is a significant heating which is significantly larger than the heating due 
to FUV in standard conditions. 

\begin{figure}
 \includegraphics[width=8cm,angle=0]{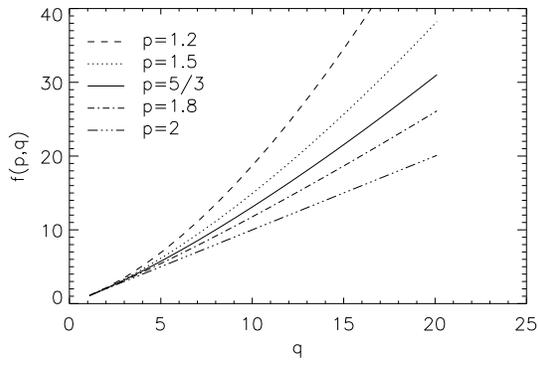}
 \caption{The numerical factor $f(p,q)$ in equation~(\ref{gam_gen}).}
 \label{fig:f}
\end{figure}

Figure \ref{fig:f} displays the numerical factor $f(p,q)$ for typical
values of $p$ and $q$. 
Note that $f(p,1) = 1$  irrespective of the value of $p$. 
For a 1-pc cloud, the typical value of $q$ is about $L/\lambda_{\rm cross} \simeq 5$.
This indicates that the actual value of $\Gamma_{\rm wave}$ is about 
 $3\sim4$ times $\Gamma_{\rm wave,0}$. 
Note that integration upto $k_{\rm max} = 2 \pi / \lambda_{\rm crit}$
 instead of $2 \pi / \lambda_{\rm cross}$, 
 would lead to $q \simeq 40$ and consequently to much larger heating 
 that would violate the condition stated by equation (\ref{Gamma_lim}).
Since equation~(\ref{gam_gen}) appears to be complex,
 two asymptotic cases are discussed in the appendix.

Figure \ref{thermax} shows the heating rate, 
 $\Gamma_{\rm wave}$ in equation~(\ref{gam_gen}) 
 for various density. For simplicity  we set $p=5/3$.
Note that we have varied the slope of the power law, $p$ 
 between 1 and 3 and found  small departures from 
 the heating obtained with the Kolmogorov spectrum as it is clear from 
 Figure~\ref{fig:f} for $q < 10$.  
 The full lines are for clouds following Larson's law so that 
 $n \propto 1/L$, while the dotted and dashed lines are for clouds of 
 gas densities 1 and 3 cm$^{-3}$, respectively. 
As expected the heating strongly depends on the cloud size $L$. 

\subsubsection{Thermal Equilibrium}
We now calculate  the thermal equilibrium for the WNM.
 We solve the energy equation 
$ n_ {WNM} \Gamma_{\rm wave}   = n_{WNM} ^2 \Lambda (T) $ where $\Lambda (T)$ is the 
cooling function described in Sect.2.3 and 
$\Gamma _{\rm wave}$ is given by  equation~(\ref{gam_gen}) for various cloud sizes, 
namely 0.25, 0.5, 1 and 2 pc. 
Figure \ref{thermal_equilibrium} shows the thermal 
 equilibrium curves.
The points which have positive slope correspond to 
 thermally stable states
 whereas those having a negative slope are unstable.  
The existence of WNM
requires that the largest pressure reached by the thermal equilibrium curve 
should be larger than the pressure inside the molecular cloud. 

As expected the largest possible density at which WNM can exist 
depends strongly on the cloud size $L$.  
For $L=1$ pc, pressures up to 10$^5$ K cm$^{-3}$  and densities 
up to $\simeq$10 cm$^{-3}$ can be reached. 
For clouds of size $L \simeq 0.25$ pc, which is close to the smallest 
wavelength $\pi \times \lambda_{\rm crit}$ allowing energy replenishment, 
pressures up to 3$\times$10$^5$ K cm$^{-3}$ and densities of about 100 cm$^{-3}$
can be obtained. We note however that, as shown in Figure~\ref{thermax}, 
for  $L < 0.5$ pc the heating is larger than $\Gamma_{\rm wave} ^{\rm lim}$ which means 
that the corresponding energy is not available in standard ISM 
conditions. 

The exact value of the highest possible WNM pressure is 
 more or less proportional to the heating which
 depends on few not-well-constrained parameters. 
In particular the amount of waves energy proportional 
 to $\langle \delta B^2 \rangle$ is a major source of uncertainties. 
However it should be noted that even with values ten times smaller, 
 the highest WNM pressure would still be a few times the standard 
 ISM pressure.

\subsection{Dissipation of Turbulent Energy}
Since  molecular clouds are turbulent it is unavoidable that turbulent
energy is continuously dissipated. Here we estimate the heating rate of the 
WNM due to the turbulent energy dissipation. We note that the dissipation of 
turbulent energy in intermittent regions of energy dissipation within
 CNM  and its consequences on the heating of the gas and the formation 
of molecules has been investigated by Joulain et al. (1998). They find that temperatures 
up  to 1000 K can be obtained in the dissipative structures.
 Padoan et al. (2000)  explore the heating by ion-neutral 
friction in three-dimensional simulations of turbulent magnetised molecular clouds and find
that this heating depends  on the position and can be much higher than 
the heating due to cosmic rays. Recently warm molecular gas 
(rotational excitation temperature of 276 K) presumably heated by the dissipation 
of turbulent energy, 
has been observed by Falgarone et al. (2005).

The mean rate of turbulent energy dissipation within 
 the whole molecular cloud is estimated to be
 $M \sigma^{3} / L$, 
 i.e., the available kinetic energy divided by 
 the crossing time of the dense material. 
It is likely that most of this energy is dissipated 
 and radiated away in the dense material, 
 and only a fraction, $\eta _{\rm turb}$, of it can be used to heat
 the warm phase. 
We therefore estimate that the warm gas within the cloud can receive an 
energy rate due to the decay of the kinematic energy of 
$\eta _{\rm turb} M \sigma^{3} / L$.
The ratio between the mass of the dense and warm components being 
 given by  $\simeq f / (1-f) \times r _{\rho}$,
 the heating rate per particle in the WNM due to kinetic 
 energy dissipation in molecular clouds is about  
 $ f / (1-f) \times \eta _{\rm turb} \Gamma _{\rm turb}$,
 where $\Gamma _{\rm turb}= r _{\rho} m \sigma ^{3} / L$.
Using the values defined in Section 2.1, we obtain:
\begin{equation}
 \Gamma _{\rm turb} = 10^{-25}  \,  {\rm erg \, s}^{-1}
 \left( { \sigma^* \over 0.4 \, {\rm km/s} }\right)^3 
 \left( { L \over 1 \, {\rm pc} }\right) ^{1/2}
\end{equation}
This indicates that for a 1-pc cloud that follows the Larson's laws, 
 the heating rate of WNM due to turbulence is small compared to 
 the heating due to mhd waves dissipation.
 $\Gamma _{\rm turb}$ is displayed in Fig. \ref{thermal_equilibrium}
 (dot-dashed thick line) which shows that
 the turbulent heating may become dominant for $L \ge 5$ pc. 
However $\eta _{\rm turb}$  is likely to be small since 
dissipation occurs within the cold component which carries the 
turbulent energy. We therefore expect that $\eta _{\rm turb} \Gamma _{turb}$
 is significantly smaller than $\Gamma_{\rm wave}$. It is therefore unclear whether the 
turbulent energy dissipation is important to heat the WNM.
It is however possible that, as proposed by Clifford \& Elmegreen (1983) and 
Falgarone \& Puget (1986), in a giant molecular cloud 
 the kinetic energy of the translational motions of cold clumps 
 could be the dominant energy. 
In this case, 
 the motion of the cold clumps excites Alfv\'en waves 
 which then cascade and dissipate providing 
  heating of the WNM at smaller scales. 

Another difficulty with the turbulent energy is that it dissipates 
 within one cloud crossing time (e.g., Mac Low \& Klessen 2004) 
 and needs to be continuously maintained by some forcing 
 which is not identified yet.
If the WNM exists deep inside molecular clouds as we propose here, 
 the high Alfv\'en velocity in the warm phase enables  
 the injection of the MHD-wave energy from the outside of 
 the molecular clouds, 
 and it might be transmitted to the cold component of 
 molecular clouds contributing to the turbulent motions. 
Since both energy densities have comparable values, 
 $E_{\rm wave} \simeq E _{\rm turb}$, 
 and since the flux of magnetic energy is larger than 
 the flux of turbulent energy, 
 $E_{\rm wave} \times V _{\rm a, WNM} \gg 
 E _{\rm turb} \times \sigma $, 
 the energy could in principle be injected inside the cloud 
 through MHD waves propagating inside WNM and 
 then partly transmitted through MHD interaction to 
 the cold component leading to near equipartition of the energies.

\section{Discussion and Conclusion}
We have investigated the possibility that WNM may exist 
 in molecular clouds
 by considering the cooling time of the WNM fluid particle and the 
 heating rate due to the mechanical energy dissipation into the cloud.
 
Our estimate indicates that the WNM when it enters into a high pressure molecular cloud
($P \simeq 10 \times P_{\rm ISM}$)
cools too rapidly to allow the existence of the warm phase inside a cloud of size 
larger than $\simeq$ 0.1 pc
unless  it is heated by some process. 
On the contrary, in a low pressure cloud ($P \simeq P_{\rm ISM}$)
the WNM can penetrate deep into a cloud of size up to $\simeq$ 1 pc.
The dissipation of the turbulent energy
 of the cold component of the molecular cloud and the photoelectric heating  
from small grains and PAH   may 
 provide enough heating only for low pressure molecular clouds 
 and unlikely for molecular clouds having pressure 10 times 
 the pressure of the ISM unless  the cloud is very turbulent
 (velocity dispersion about 2 times the value of the  Larson's law)
or located near a strong UV sources ($G_0 \simeq 10$).

In contrast, the dissipation of the MHD waves seems to be a promising
 mechanism to maintain warm atomic gas inside high pressure molecular clouds.
The value of the highest pressure at which WNM may exist depends 
on the cloud size, density, magnetic intensity and ionization degree.
We find that for a 1-pc cloud, WNM may exist up to pressures of about 
$10 \times P_{\rm ISM}$.

The main difference with the previous estimate for 
 the case of the standard ISM (e.g. Ferri\`ere et al. 1988)
 is that, since the ionization is about 10 times smaller than 
 in the standard ISM, 
 the cooling rate is few times smaller, 
 whereas the heating rate is  larger because the dissipation time is smaller.
Also  the magnetic field is stronger and the wavelengths smaller.

Finally we suggest that if channels of warm diffuse and magnetised gas 
do exist and permeate the molecular clouds, then the dissipated energy 
can be replenished from the outside because of the high Alfv\'en 
velocity in the WNM. We speculate that such channels
may also help to inject energy continuously into the dense component of 
the molecular cloud helping to sustain the turbulent motions 
 that otherwise decay in a crossing time 
 (e.g., Mac Low \& Klessen 2004).

\begin{acknowledgements}
This paper was written while PH was a visitor at  Kyoto University,
and he is very grateful for the support and hospitality he received 
during his visit. 
We thank an anonymous referee for his comments and interesting suggestions 
which have been helpful to clarify the paper.
We acknowledge a critical reading of the manuscript by
 David Hollenbach as well as stimulating discussions.
PH is grateful to Maryvonne G\'erin, Katia Ferri\`ere 
and Sylvie Cabrit for very helpful  discussions on various physical processes
relevant for this work. He is most grateful to Edith Falgarone for 
extensive discussions on the magnetically interacting clump model
of Falgarone \& Puget (1986).
This work is supported by the Grant-in-Aid for the 21st Century COE 
 ``Center for Diversity and Universality in Physics'' from 
 MEXT of Japan.
\end{acknowledgements}

\appendix

\section{Asymptotic Limits for Heating due to MHD Wave Dissipation}
Since equation~(\ref{gam_gen}) appears to be complex two asymptotic cases 
are discussed.

The first asymptotic case is when the size of the cloud,
 $L$, is close to  
 $\lambda_{\rm min}= 2 \pi \times \lambda_{\rm crit}$ then 
 $k _{\rm max } \simeq  2 \pi / \lambda _{\rm min}$  and
 $k _{\rm max } = \alpha k_ {\rm min }$ where 
 $\alpha $ is a numerical factor of a few, and we have
\begin{equation}
\Gamma_{\rm wave} = \Gamma _{\rm wave} ^{{\rm max}} \times
 {1 \over 2  \alpha  ^2} 
 { \alpha ^{4/3} -1  \over 1 - (\alpha)^{-2/3}},
\label{gam_alp}
\end{equation}
where $  \Gamma _{\rm wave} ^{{\rm max}} = \Gamma _{\rm wave,0} (\lambda=\lambda_{\rm min}) $
and is given by 
\begin{equation}
  \Gamma _{\rm wave} ^{{\rm max}} 
  \simeq  4 \times 10^{-23} \, {\rm erg \; s^{-1}}
 \left( {  n_{\WNM} \over 1 \, {\rm cm}^{-3} } \right) ^{1/2}
\left( {T \over 8000 {\rm K} } \right)^{0.725}
\left( \zeta \over 3 \times 10^{-17} {\rm s}^{-1} \right)^{1/2}
\left( {B^* \over 10 \, {\rm \mu G} } \right)^2.
\label{gammax}
\end{equation}
Since $\Gamma _{\rm wave,0}$ decreases with $\lambda$,   $\Gamma _{\rm wave} ^{{\rm max}} $
is an upper limit for the heating rate due to MHD dissipation.
Unlike $\Gamma _{\rm wave} ^ 0 $,  $  \Gamma _{\rm wave} ^{{\rm max}} $ increases
with the cosmic rays ionization rate and  
with the gas density. This is due to the fact that $\lambda_{\rm min}$ decreases 
with these two parameters and that the dissipation time is $\propto \lambda^{-2}$.
For $\alpha=2$, we have $\Gamma_{\rm wave} \simeq 
0.5 \times \Gamma _{\rm wave} ^{\rm max}$. 

The second asymptotic case that we consider is when 
 the size of the cloud is much larger than $\lambda_{\rm min}$,
$k _{\rm max } = 2 \pi / \lambda_{\rm cross} \gg  k_ {\rm min } = 2 \pi / L$.  
Then we can write
\begin{equation}
\Gamma_{\rm wave} \simeq
 \Gamma _{\rm wave,0} (\lambda=\lambda_{\rm cross})  \times
 {1 \over 2} \left( {    \lambda_{\rm cross}} \over L \right) ^{2/3} .
\label{gam_inf}
\end{equation}
Combining this equation with equation \ref{lcross}, we obtain
\begin{equation}
  \Gamma _{\rm wave}  
  \simeq 1.6 \cdot 10^{-24} {\rm erg \; s^{-1}}
 \left( {   n_{\WNM} \over 1 \, {\rm cm}^{-3} } \right) ^{-1/6}
\left( {T \over 8000 {\rm K} } \right)^{-0.242}
\left( \zeta \over 3 \cdot 10^{-17} {\rm s}^{-1} \right)^{-1/6}
 \left( {L \over 1 \, {\rm pc}} \right) ^{-4/3} 
\left( {B^* \over 10 \, {\rm \mu G} } \right)^{10/3}.
\label{gam_ps}
\end{equation}
This expression depends only weakly on the gas density and ionization rate. It
decreases with the size of the cloud, $L$, less rapidly than in the monochromatic 
case stated by equation~(\ref{gamwave}). Note that for $k_{\rm min}= 2 \pi / \lambda _{\rm min}$
instead of $k_{\rm min}= 2 \pi / \lambda _{\rm cross}$, one obtains 
$\Gamma _{\rm wave} \propto  n_{\WNM} ^{1/6} \zeta^{1/6} L_c^{-2/3}$, indicating 
an even  shallower dependence on the cloud size.

\clearpage

\clearpage

\clearpage

\clearpage

\clearpage

\end{document}